\documentclass[twocolumn,prl,showpacs]{revtex4}

\usepackage{graphicx,amssymb,amsmath}
\usepackage[applemac]{inputenc}

\begin{document}

\title{Non stationary wave turbulence in an elastic plate}
\author{Benjamin Miquel and Nicolas Mordant}
\email[]{nmordant@ens.fr}
\affiliation{Laboratoire de Physique Statistique, Ecole Normale Sup\'erieure \& CNRS, 24 rue Lhomond, 75005 Paris, France.}


\begin{abstract}
We report experimental results on the decay of wave turbulence in an elastic plate obtained by stopping the forcing from a stationary turbulent state. In the stationary case, the forcing is seen to induce some anisotropy and a spectrum in disagreement with the weak turbulence theory. After stopping the forcing, almost perfect isotropy is restored. The decay of energy is self similar and the observed decaying spectrum is in better agreement with the prediction of the weak turbulence theory. The dissipative part of the spectrum is partially consistent with the theoretical prediction based on previous work by Kolmakov. This suggest that the non agreement with the weak turbulence theory is mostly due to a spurious effect of the forcing related to the finite size of the system.
\end{abstract}

\maketitle
Wave turbulence is observed in systems involving a large number of waves coupled by nonlinear effects. Like hydrodynamical turbulence, it is characterized by wide ranges of excited length scales and time scales.  As wave turbulence can be weakly nonlinear, a statistical theory (Weak Turbulence Theory, WTT) could be analytically derived and applied to numerous systems such as optics, oceanic/atmospheric waves or plasmas~(see~\cite{Newell1,Zakharov} for reviews). The predicted phenomenology of the forced case is very similar to that of hydrodynamical turbulence: the energy injected at large scales where dissipation is negligible is transferred conservatively to small scales by nonlinearity through the Kolmogorov-Zakharov (KZ) cascade. The cascade operates until the energy reaches scales that are small enough so that dissipation dominates over non linear transfers. In this respect wave turbulence can appear as an intermediate step in complexity towards the understanding of hydrodynamical turbulence. 

Assuming that the waves composing the motion exchange energy at a slow rate compared to their frequency (the non linear interactions are weak) and that the system is infinite, the WTT yields analytical predictions for spectra and other other statistical quantities. Let $a_{\mathbf k}$ stand for the canonical variables that diagonalize the linear part of the Hamiltonian of the motion. The wave action (or occupation number at wavevector $\mathbf k$) is defined as $n_{\mathbf k}=\langle a_{\mathbf k}(t)a_{\mathbf k}^\star(t)\rangle$, where $\langle\,\rangle$ is a statistical average on realizations and $^\star$ stands for complex conjugaison (see \cite{During} for details in the case of flexion waves on a thin elastic plate). In the framework of the WTT, the time evolution of $n_{\mathbf k}$ can be described  by
\begin{equation}
\frac{\partial n_{\mathbf k}}{\partial t}=\mathcal Col(n_{\mathbf k})-\gamma_k n_{\mathbf k}+F_{\mathbf k}(t)
\label{KE}
\end{equation}
The forcing $F_{\mathbf k}$ is usually acting on large scales and the dissipation $\gamma_{\mathbf k}n_{\mathbf k}$ is efficient at small scales. In the intermediate inertial range of scales, energy is transferred conservatively in Fourier space. $\mathcal Col(n_{\mathbf k})$ is the collision integral whose expression was derived in many physical systems~\cite{Newell1,Zakharov,During}. It appears as the divergence of an energy flux in Fourier space and formally  it is reminiscent of the kinetic theory of gases as the energy is transferred nonlinearly by ``collisions" of resonant wavetrains~\cite{Newell1,Zakharov}. The Kolmogorov-Zakharov (KZ) spectrum is a stationary solution of the forced case whose analytical expression can be derived in the intermediate range of scales. 
%The KZ spectrum cancels the collision integral. In the case of a thin elastic plate, the KZ spectrum was derived in~\cite{During}. 

Here we consider the case of flexion waves in an elastic plate. In the experiment we measure the velocity spectrum $E(\mathbf k)$ which is related to $n_{\mathbf k}$ by $E(\mathbf k)\propto k^2 n_{\mathbf k}$\cite{During}. The theoretical prediction for the velocity spectrum is:
\begin{equation}
E(\mathbf k)=CP^{1/3}\ln^{1/3}({k_c}/{k})
\label{theo}
\end{equation}
where $C$ is a dimensional constant that can in principle be calculated analytically, $P$ is the average input power and $k_c$ is a cutoff frequency that has to be introduced for the sake of dimensionality . The logarithmic dependency has to be introduced to ensure a finite flux of energy because of the degeneracy with the equipartition spectrum $E(\mathbf k)=constant$ which has a zero flux of energy~\cite{During}. The relation between the cutoff wavenumber $k_c$ and the small (or large) scale dissipation is largely an open question as no dissipation is considered in the theory. Although D\"uring {\it et al.} claim to have observed the KZ spectrum in numerical simulations (by choosing $k_c$ as the dissipation scale of the numerical simulation) so far the KZ spectrum remains elusive in experiments as the theoretical scaling in $P$ and in $k$ (or $\omega$) is not observed~\cite{Boudaoud,epjb}. More generally, the applicability of the Weak Turbulence Theory to real systems remains an open question and only few experimental systems are available for a quantitative comparison with the theoretical predictions (see the recent review~\cite{Newell1} for a detailed discussion). In this respect, the case of flexion waves in a thin plate is promising because the use of a high speed profilometry technique enables advanced measurements of the full space-time structure of the turbulent field~\cite{Cobelli,epjb,Mordant}. Although not in agreement with the WTT, the flexion wave turbulence has been observed to be indeed weakly non linear, with a persistence of waves and a non linear dispersion relation weakly modified from the linear dispersion relation. These observations are consistent with the requirement and the predictions of the WTT. It has been suggested that either dissipation or finite size effects could be responsible for such discrepancy~\cite{Boudaoud,Cobelli,epjb,Mordant}. Here we study a non stationary regime of such turbulence obtained by stopping the large scale forcing after a stationary regime is reached. By observing the decline of turbulence, we suggest that the forcing itself is responsible for the disagreement between the measured spectrum and the theoretical one through finite size effects.

Non stationary weak turbulence was studied theoretically or numerically by several authors~\cite{Zakharov,Falkovich,Connaughton1,Connaughton2} but not in conditions that are relevant to laboratory studies. Usually dissipation is not considered and energy is initially localized in a finite interval of (large) scales so that the propagation of an energy front in Fourier space to large wavenumbers was observed~\cite{Connaughton1,Connaughton2}. The experimental situation of decaying wave turbulence where dissipation is present and starting from stationary turbulence is different and was studied experimentally for capillary wave turbulence~\cite{Kolmakov1} but with limited measurements due to the difficult experimental conditions. A self similar decay of the spectra was predicted in this case~\cite{Kolmakov2} but has not been observed yet.

\begin{figure}[!htb]
\includegraphics[width=7.5cm]{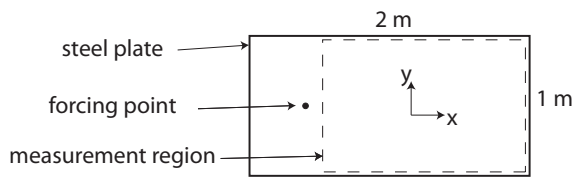}
\caption{Schematics of the experimental setup (similar to~\cite{epjb}). A $2\times1$~m$^2$ plate, $0.4$~mm thick, is vibrated by an electromagnetic shaker at 30~Hz. A single value of the forcing amplitude is studied here. The deformation of the plate is measured by a high speed profilometry technique~\cite{Cobelli1} over a surface $1.25\times0.94$~m$^2$. Linear waves follow a dispersion relation $\omega=ck^2$ with $c$ close to 0.64~m$^2$.s$^{-1}$ so that the 30~Hz forcing corresponds to a wavelength $\lambda\approx 0.37$~m. Movies of the deformations were recorded at 6000~frames/s. }
\label{set}
\end{figure}
The forcing is stopped after the plate reached a stationary turbulent state. Averages over realizations  are performed by repeating the experiment 90 times (see fig.~\ref{set} for a sketch of the experimental setup). Movies of the deformation of the plate are obtained by high speed profilometry~\cite{Cobelli}. Fourier analysis of the data is similar to~\cite{epjb}. At a given $\mathbf k$, two counter propagating waves are present with frequencies of opposite signs so that the spatial Fourier transform of the velocity field can be written generally as 
$v(\mathbf{k},t)=v^+(\mathbf{k},t)+v^-(\mathbf{k},t)$ where $v^+$ (resp. $v^-$) contains only positive (resp. negative) frequencies. We study only Fourier components of negative frequencies (i.e. $v^-$) by applying a discrete Hilbert transform in time (without loosing any information as $v^+(\mathbf k,t)=v^{-\star}(-\mathbf k,t)$). In this way, we can separate waves at a given $\mathbf k$ traveling in the two opposite directions so that to study isotropy issues. We chose to perform the statistical analysis on $v^-$ so that 
for instance waves with $k_x>0$ are propagating in the direction $x>0$ which makes the discussion slightly easier. Here we define the 2D time dependent Fourier spectrum as $E(\mathbf k,t)=\langle | v^-(\mathbf k,t)|^2\rangle$. Note that this spectrum is no longer symmetrical under the change $\mathbf k\rightarrow -\mathbf k$. $E(k=\| \mathbf k\| ,t)$ is the angle averaged spectrum $E(k,t)=\int_0^{2\pi} E(\mathbf k,t)kd\theta$ where $\theta$ is the polar angle of $\mathbf k$. To get some insight on anisotropy issues, the interval of integration can be restricted to four quadrants of angles in $[-\pi/4,\pi/4]$, $[\pi/4,3\pi/4]$, $[3\pi/4,5\pi/4]$ and $[5\pi/4,7\pi/4]$. These intervals correspond to propagation along directions toward $x>0$, $y>0$, $x<0$, $y<0$ respectively (noted $x^+$, $y^+$, $x^-$, $y^-$ resp. in the following).

\begin{figure}[!htb]
\centering
(a)\includegraphics[width=8cm]{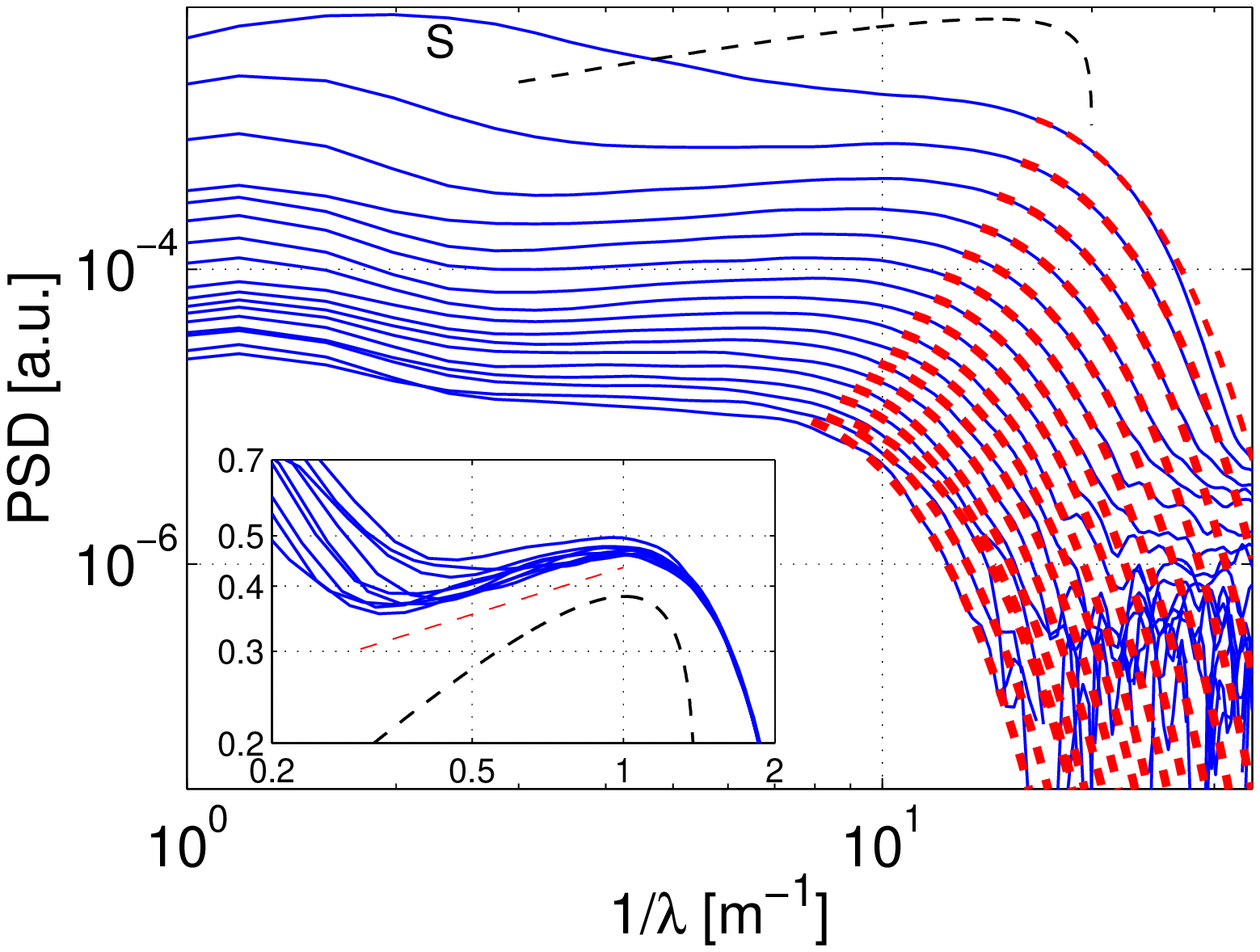}
\includegraphics[width=8.5cm]{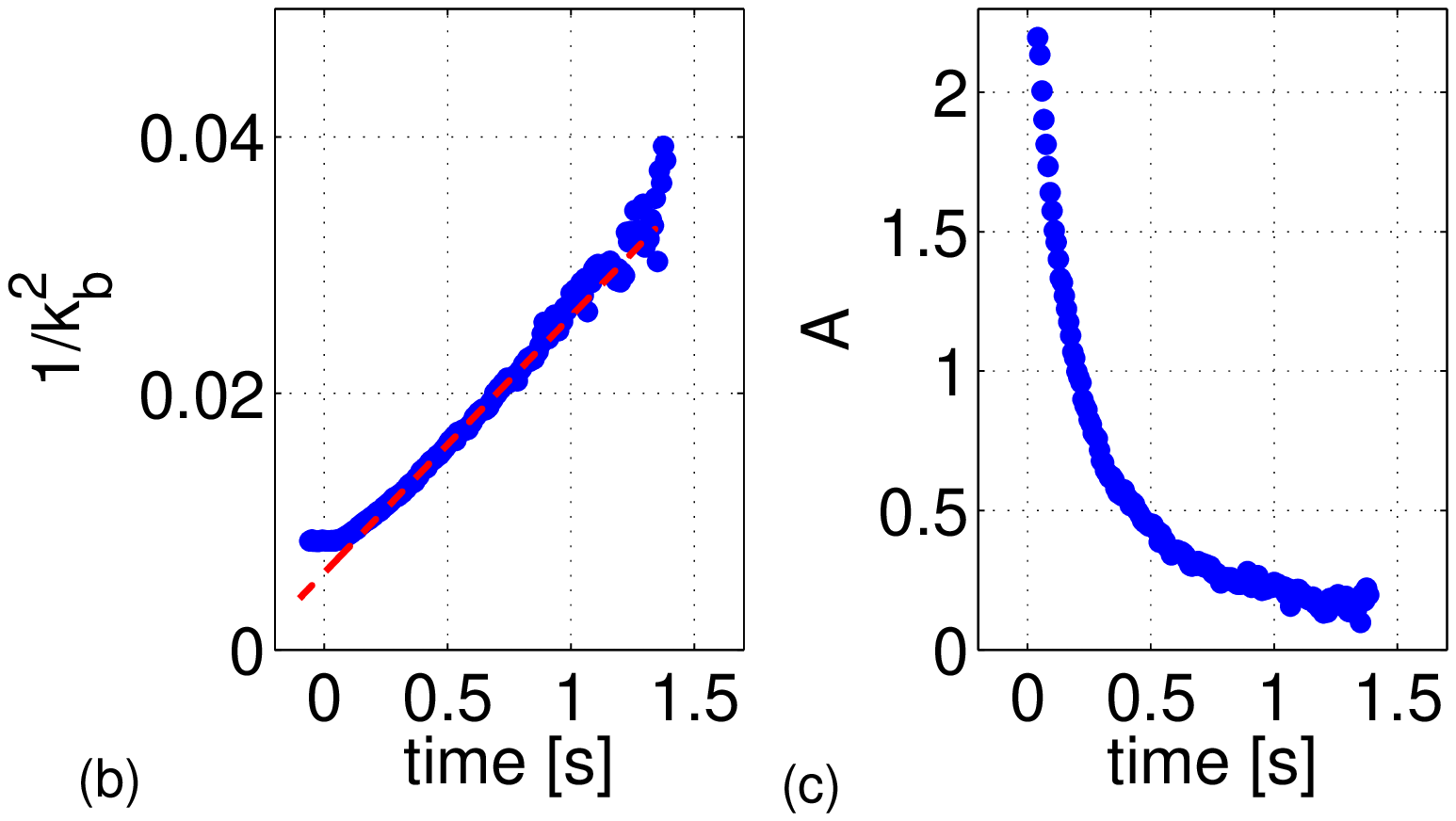}
\caption{(a) Decay of the velocity spectrum $E(k=2\pi/\lambda,t)$ at various times. Top solid curve (S): stationary spectrum. Lower solid curves, from top to bottom: 104~ms after stopping the forcing and every 83~ms subsequently (spectra are averaged over 42~ms). Thin dashed line at the top of the figure: shape of the theoretical spectrum ($E(k)\propto k\ln^{1/3}(k_c/k)$). Thick dashed line: fit of the model (\ref{diss}) for the dissipative part of the spectrum at each time. 
{\it Insert}: rescaling of the decaying spectra using the fitting parameter of (\ref{diss}) at each time as a function of $k/k_b(t)$ (covering both the inertial and the dissipative range) for times between 0.1 and 0.83~s. Upper dashed line: power law $k^{0.3}$, lower dashed line: theoretical spectrum. 
(b) fitting parameter $B/k_b^2(t)$ as a function of time. Dashed line: linear trend. (c) fitting parameter $A$ as a function of time.}
\label{dec}
\end{figure}
An example of the decay of the velocity spectrum is shown in fig.~\ref{dec}(a). The top solid curve shows the stationary spectrum observed before stopping the forcing shaker. A maximum is observed at the forcing wavenumber and a short power law inertial regime is observed with an exponent about $-0.2$ corresponding to the previously observed $\omega^{-0.6}$ scaling of the single point frequency spectrum (change of variable $\omega\leftrightarrow k$ is obtain through the dispersion relation $\omega\propto k^2$)~\cite{Boudaoud,Mordant}. The theoretical spectrum (\ref{theo}) is shown with a thick dashed line and is clearly different from the experimental one both in the fact that the latter is increasing in the inertial range and that the cutoff is not following a logarithmic decay. After stopping the forcing and after a short transient during which the forcing peak disappears, the spectra are decaying in a self similar way until the inertial range is suppressed. A peak at low $k$ is observed which corresponds to the energy of the lowest frequency modes of the plate which are not resolved by our measurement (the field of view being too small). The cutoff wavenumber decays with time. The exponent of the inertial range scaling is observed to be close to 0.3 i.e. closer to the theoretical exponent (equal to one if one discards the log correction in $E(k)\propto k\ln^{1/3}(k_c/k)$) than for the forced case ($-0.2$). The remaining disagreement may be attributed to a too narrow interval of length scale available for the energy cascade or to the logarithmic corrections.

\begin{figure}[!htb]
\centering
\includegraphics[width=8cm]{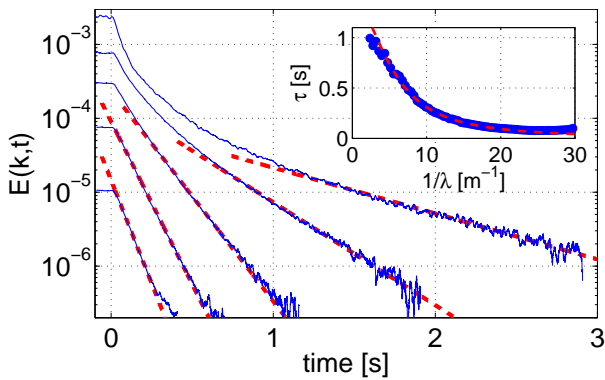}
\caption{Decay of energy. Main figure: evolution of the angle averaged power spectrum $E(k,t)$ for $1/\lambda=k/2\pi=5$, $10$, $15$, $20$, $25$~m$^{-1}$ from top to bottom (semilog scale, the curves have been shifted vertically for clarity). Dashed lines: eye guides for the long time exponential decay of the energy. Insert: evolution of exponential decay times as a function of $k$. Dashed line: Lorentzian fit $(0.73+0.025(k/2\pi)^2)^{-1}$.}
\label{dec2}
\end{figure}
The time evolution of the energy at a given $k$ is shown in figure~\ref{dec2} for a few values of $k$. The decay occurs in two steps: the first one resembles a stretched exponential shape, the second one is exponentially decaying. 
%As the high $k$ cutoff is drifting with time to lower values, a given $k$ initially in the inertial part of the spectrum in fig.~\ref{dec1}(d) ends up in the fast decaying region of the spectrum. 
The change of behavior in the decay is related to a change in dynamics from the non linear cascade to the dissipative decay. The characteristic times of the exponential decay are shown in the insert of fig.~\ref{dec2}. For $k$ above the forcing scale, they tend to follow a Lorentzian variation with $k$ suggesting the coexistence of two processes for dissipating energy at large or small scale (the increase at the upper values of $k$ is due to a very poor signal over noise ratio). The values of the dissipative times (from 1~s to $1/10~s$) show also a clear scale separation from the period of the wave: the ratio equals 25 at the forcing scale up to over 100 at $k/2\pi=25$~m$^{-1}$. The exponential decay is consistent with the dissipative term $-\gamma_k n_{\mathbf k}$ with $\gamma_k=0.73+0.025(k/2\pi)^2$~s$^{-1}$. The observed time scale separation opens the possibility of an energy cascade as predicted by WTT.

A theoretical prediction for the shape of the decaying spectrum can be derived following the theoretical work of Kolmakov~\cite{Kolmakov2}: one looks for a self similar decaying solution of (\ref{KE}) expressed as $n_k(t)=Ak_b(t)^{\alpha}g(\xi)$ where $k_b(t)$ is a time dependent dissipative cutoff wavenumber and $\xi=k/k_b(t)$. We assume that $\gamma_{\mathbf k}\propto k^2$ as suggested by the data. In order to express (\ref{KE}) as a function of the single parameter $\xi$ for a self similar decay of the spectrum, the scaling properties in $k$ of the collision integral (as given in \cite{During}) and of the dissipation time impose that $\alpha=0$ in the case of the plate and the cutoff wavenumber must verify $\dot{k}_b(t) k_b(t)^{-3}=constant$. One gets $k_b^2(t)=k_b^2(0)/(1+t/\tau)$ where $\tau$ is a constant characteristic time.
In the dissipative range $k\gg k_b(t)$, the collision integral in (\ref{KE}) is negligible compared to the dissipative term so that (\ref{KE}) yields $g'(\xi)=-2B\xi g(\xi)$ (where the prime stands for derivation and $B$ is a positive constant) so that $g(\xi)\propto \exp(-B\xi^2)$. The overall dissipative region of the spectrum is thus expected to follow 
\begin{equation}
E_d(k,t)=A k^3\exp (-Bk^2/k_b^2(t))
\label{diss}
\end{equation}
where $A$ is a positive constant depending on the spectrum at initial time.
The dissipative cutoff is then expected to follow a Gaussian decay with $1/k_b^2(t)$ being affine with time. The decay (\ref{diss}) has been fitted to the dissipative region in fig.~\ref{dec} that provides $A$ and $B/k_b^2(t)$ (fig.~\ref{dec} (b) and (c)). The agreement is fairly good and the cutoff wavenumber follows the theoretical prediction ($1/k_b^2(t)$ being affine with $t$). Nevertheless the prefactor $A$ of the exponential decay is not constant as predicted above but it is strongly decaying with time. Thus, the shape of the spectrum is preserved but a genuine self-similarity is not observed. It is interesting to notice that the prediction of $k_b(t)$ and the dissipative behavior of the spectrum actually depend on the scaling properties of the collision integral $C(n_{\mathbf k})$ and that they are at least partially compatible with the data. Note that the cutoff observed in the stationary forced regime is also close to such a Gaussian decay. In the inertial range ($k<k_b(t)$), the collision integral dominates and the spectrum is expected to follow the KZ spectrum~\cite{Kolmakov2}.

\begin{figure}[!htb]
\centering
\includegraphics[width=9cm]{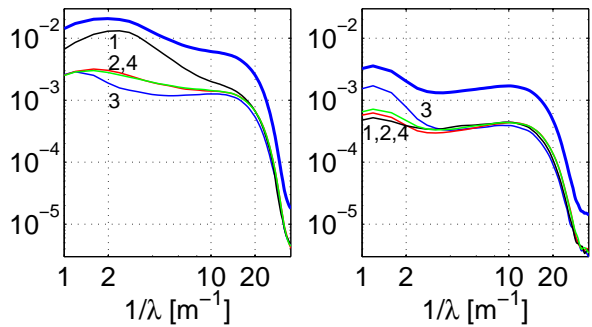}
\caption{Check of isotropy of the spectrum. (a) stationary forced regime. (b) decaying case after $0.18$~s of decay. The upper thick curve is the fully angle averaged spectrum. The 4 lower curves correspond to an angle average over the 4 quadrants (see text).  Curve 1 corresponds to propagation directions $x^+$, curve 3 to $x^-$ and 2 and 4 to both directions along $y$ (almost superimposed).}
\label{dec1}
\end{figure}
The issue of isotropy is addressed by separating the fully angle averaged spectrum $E(k,t)$ into the four quadrants defined above (as shown in fig.~\ref{dec1}). In the forced case, waves in the quadrant $x^+$ dominates strongly. Due to the position of the measurement region, the spherical waves directly generated by the shaker are only partially visible and lie mostly in this $x^+$ quadrant. This explains that the energy in this quadrant is dominating over the other quadrants at the forcing wave number and most of the inertial range. Waves propagating in the other quadrants have been generated by nonlinearities and/or rebounds on the boundaries of the plate. The forcing is only weakly visible on the curves 2 and 4 ($y^+$ and $y^-$ quadrants ) and not visible in the $x^-$ quadrant . In contrast to the forced case, the decaying spectrum is almost perfectly isotropic. Note that the peak at the lowest $k$ corresponds to energy in the lowest frequency modes with wavelengths larger than the field of view of our technique. Some energy is stored in these modes although their frequency is lower than that of the forcing. Their energy is also decaying but may act as an energy reservoir that feeds the cascade during the decay. It should be noted that the spectrum of the $x^-$ quadrant  is not much different in the forced and the freely decaying regimes.

We interpret our observations as follows: In the weakly forced regime, non linear effects are due to cumulative effects over several periods of the waves. Due to the propagating nature of the waves, it translates into some propagation length required for the generation of other frequencies by weak non linearities. If this distance is not short compared to the size of the plate, the transitory regime dominates: this is the case in the $x^+$ quadrant. By contrast the waves in the $x^-$ quadrant have propagated further (at least one rebound) and this seems enough for non linear effects to have transferred energy from the forcing to other waves. This explains why the spectrum of the waves propagating in this direction are close to that of freely decaying turbulence. 
We expect that a bigger system and a measurement region further from the forcing would yield a better agreement with the theory. Thus our observation can be assigned to finite size effect, the measurement region being too close to the forcing point. It is very different from another kind of finite size effect related to the quantization of the plate modes due to its size. The latter effect is expected to influence the energy cascade by limiting the number of wave resonances available to transfer energy~\cite{Kartashova}.

The exponent of the isotropic decaying spectrum is closer to the WTT prediction of the Kolmogorov-Zakharov spectrum but not equal. The remaining discrepancy can be attributed either to a too short inertial range and/or to the logarithmic correction predicted by WTT. The self similar decay implies that the logarithmic cutoff $k_c$ should evolve in time following the decay of $k_b(t)$ and thus display a dissipative scaling (as implicitely assumed in the numerical simulations of~\cite{During}). The partial agreement of the observed and predicted dissipative cutoff provides some support for the validity of the expression of the collision integral derived by the WTT. The fact that the prefactor $A$ is not constant in time is most likely due to the fact that the damping coefficient $\gamma_k$ is not purely quadratic in $k$. The constant part of $\gamma_k$ may accelerate the decay of energy as dissipation occurs at all wavenumbers. This may be responsible for the observed decay of $A$.

\begin{acknowledgments}
This work was funded by the French Agence Nationale de la Recherche under grant TURBONDE BLAN07-3-197846.
\end{acknowledgments}

\bibliography{plaqueiv}

\clearpage

 \end{document}